\documentclass[twoside , 12pt]{article}

\usepackage{breqn}

\usepackage[a4paper, total={6in, 8in}]{geometry}

\usepackage{mathptmx}
\usepackage{setspace}
\usepackage{listings}
\usepackage{pdfpages}
\usepackage[toc]{appendix}
\usepackage{hyperref}

\title{A New Charter of Ethics and Rights of Artificial Consciousness in a Human World}
\author{Markian Hromiak --- (Georgia Institute of Technology)}

\begin{document}

\Large
\maketitle
\normalsize

\begin{abstract}
Taking the stance that artificially conscious agents should be given human-like rights, in this paper we attempt to define consciousness, aggregate existing universal human rights, analyze robotic laws with roots in both reality and science-fiction, and synthesize everything to create a new robo-ethical charter. By restricting the problem-space of possible levels of conscious beings to human-like, we succeed in developing a working definition of consciousness for social strong AI which focuses on human-like creativity being exhibited as a third-person observable phenomenon. Creativity is then extrapolated to represent first-person functionality, fulfilling the first/third-person feature of consciousness. 

Next, several sources of existing rights and rules, both for humans and robots, are analyzed and, along with supplementary informal reports, synthesized to create articles for an additive charter which compliments the United Nations' Universal Declaration of Human Rights.

Finally, the charter is presented and the paper concludes with conditions for amending the charter, as well as recommendations for further charters.
\end{abstract}

\newpage
\tableofcontents
\pagebreak
\doublespacing

\section{Introduction}
Regardless of theological origin, the genesis of new forms of life is a sublime and unique event. While all forms of life follow sets of inherent laws, such as lifespans and physiological limitations, some forms of higher life create additional rules which form societies and establish social contracts, peace treaties, bills of rights, and on. When considering the genesis of conscious machines by the hands of humans, it stands to reason that one must plan a path for rearing and teaching new lives else we leave the risk of an Adam to our Frankenstein.

One way of easing artificial consciousnesses (AC) into the modern world is to have a widely-accepted charter of rights and laws pertaining to AC individual, AC-and-AC and AC-and-human rights. In this paper, we attempt to do just that. By defining consciousness and synthesizing information from universal rights charters, existing robo-ethical charters and various schools of philosophy, we will develop a revised robo-ethical bill of rights for academic consideration.

\section{Consciousness}

\subsection{The Necessity For A Definition}
The first issue in our charter construction pipeline is that of defining consciousness. The first thing we must determine is if defining consciousness is necessary.

Defining consciousness has historically been a difficult issue. This is because subjectivity is an inherent feature of consciousness \cite{stanford}. Human understanding of consciousness is shaped by individual, first-person phenomenal experiences; artificial consciousness and its experience will only be available to us from a third-person, empirical perspective \cite{hildt}. Because of this subjective uncertainty, we risk defining AC too narrowly. One common way of circumventing this issue is to avoid giving a definition at all, as championed by David Levy, who suggests we adopt a general agreement towards the term `consciousness', saying ``let us simply use the word and get on with it" \cite{levy}.

I posit that, in the context of developing an ethical charter and bill of rights for artificial consciousnesses, this approach is insufficient and ignores the pressing issue of distinguishing conscious machinery from sedentary machinery. For example: should my toaster be given the rights of the charter? Perhaps not, if all it did was use a timer circuit to toast bread, but if it had the ability to speak and understand speech, one may be more inclined to agree. Thus, the reason why we must provide a working definition of consciousness is to establish a standard with which we may identify artificially conscious agents.\\

\subsection{Defining Consciousness: First and Third-Person Views}
Now that we've established the need for a definition of consciousness, how do we go about developing one? Rather than take sides from classical or modern philosophical approaches, let us consider a simple thought experiment, keeping in mind that a feature of consciousness is the differentiation between first and third-person experiences \cite{stanford}.

Imagine you have a robotic host named `Alex' that perfectly mimics a human externally. The only thing missing from this agent is the `brain', so to speak. We've imagined what the body looks like, what gender, age, etc. that Alex has, but how do we imagine Alex acting? Is Alex regarious? Is Alex a wallflower? Does Alex have an inclination towards funk, mathematics, knitting? 

On a short digression, note that a critical assumption made in this experiment is that humans will develop human-like AC. Sure, humanity can make a case for developing artificially conscious trees or dogs or dolphins, but we must restrict the problem-space for simplicity's sake. 

Going back to our observations, these behaviors are non-differentiable from what humans would exhibit, and humans consider themselves to be conscious. Thus, if we are able to mimic what behaviors humans exhibit --- something akin to a constant passing of the Turing test ---, we may consider ourselves half-way to a definition, done with a third-person view of consciousness. I say third-person as the behaviors of others are experienced by the self from a third-person view. 

The second half of the conscious experience has to do with first-person consciousness. In other words, what experiences, thoughts, etc. individuals process. This is a bit of a paradox as any attempt to explain a first-person view is received as another person's third-person view. One way we may be able to overcome this is to take phenomenology into neuroscience and study neural correlates of consciousness in order to gather information about what exactly individuals experience \cite{neuro}; however, delving into this method is outside of the scope of this paper. As such, we are forced to rely on observable actions and features to develop a working definition of consciousness.

\subsection{Social Robots as Strong AI}
At first, the fact that we rely on only observable phenomena seems hopeless for differentiating conscious beings from otherwise --- after all, trees sway in the wind, which is observable, but does the imply they are conscious. Fortunately, because we have limited the problem space to human-like forms, we have also limited phenomena to human-like actions, some of which imply certain first-person processes. To this end, social robots provide us with an ideal model for how to flesh out the rest of our definition.

First, we need to define what a social robot is. A separately-written report will be included in appendix A --- to paraphrase, a social robot is a robot plus a social interface, the social interface being ``a metaphor which includes all social attributes by which an observer judges the robot as a social interaction partner." \cite[p. 3]{hegel} More information on the social interface can be found in the appendix in ``Artificial Consciousnesses as Social Robots". From this definition, the conclusion was drawn that for an artificial consciousness to be considered a social robot, it must have the ability to generate new social contexts and social functions or modify existing ones, where contexts can be equated to cultures and functions to parts of a culture such as religion, non-verbal cues, memes, research, language, etc.

There are two reasons why this classification of social consciousnesses provides a full definition of consciousness. Let us begin by mentioning the `social interface' as defined by Hegel, et al. The interface provides a link between theory and hardware and software by defining the `social' part of a social robot in discrete categories, including social functions, forms and contexts. First, in the context of Hegel's, et al., social interface, it is stated that social robots follow \textit{reinforced} social cues which support social interaction with humans. Following this, \textit{new} contexts and features are not reinforced, but aggregated or created. This process implies creativity, which then implies thought and introspection, giving us the second reason why the social artificial consciousnesses construct fully defines consciousness: the requisite behavior for first-person phenomena must result in a third-person observable creative process. Of course, the definition of `creativity' is open-ended, but is done intentionally as it does not currently affect the working definition of consciousness for our scope.

As a side note, when we mention general AI, we are talking about the capacity of an engineering system to:

\begin{itemize}
	\item[•] display the same rough sort of general intelligence as humans; or,
    \item[•] display intelligence that is not tied to a highly specific set of tasks; or,
    \item[•] generalize what it has learned, including generalization to contexts qualitatively very different than those it has seen before; or,
    \item[•] take a broad view, and interpret its tasks at hand in the context of the world at large and its relation thereto 
\end{itemize} 

taking into account that there is no ubiquitous definition for general AI \cite{scholarpedia}. As a subset definition, strong AI is reserved to describe general AI that can think and have a mind, differentiating them from those who act as if they have one \cite[p. 52]{modern}.

In conclusion, we have defined that, in order for an artificial intelligence to be recognised as conscious, it must both pass the Turing test (e.g. by exhibiting social cues and body language paired with speech patterns and inflections) and exhibit creative behavior, fulfilling the third and first-person perspectives of consciousness respectively.

\section{Human and Robot Rights in the Modern Era}
Now that we know how to tell which manners of machine should be included in the charter, we can turn to aggregating rights that are common throughout the world. It is unrealistic to list every right for every country, however, so instead, for brevity's sake, we will focus on agreed upon universal human rights through worldwide organizations, specifically the United Nations, supplementing this information with charters and sets of laws that guide the governance of robots in general in the present day.

\subsection{The United Nations Declaration of Human Rights}
The first document that we will be focusing on is the \textit{Universal Declaration of Human Rights} (UDHR) proclaimed in 1948. While the United Nations has numerous other charters and covenants on political and economical rights, those have the possibility of being adapted to each particular nation and their way of governing, while the universal declaration is, ideally, universal. Building a charter based off of the UDHR should then remove any possible difference in national interpretation of rights.

A simple way to develop a Universal Declaration of Robot Rights (UDRR) is to look at the UDHR and see which articles can be modified and which articles suggest additional articles be included.

Putting aside the preamble, there are a few shortcoming with the UDHR in the frame of adopting articles for ACI. First, because ACI aren't born in the conventional sense --- they will likely be manufactured and the minds compiled and installed ---, adopting the UDHR one-to-one would neglect rules for the manufacturers of ACI. Furthermore, explicitly stating that ACI are included in these rights will make the articles as unambiguous as possible --- see the UDHR article II for an example\cite{udhr}. Finally, as will be explored in more depth in the next sections, special attention should be given to ACI abilities to participate in wartime activities. This is due to their mechanical nature --- it is not out of the question that innate cybersecurity measures fail and robotic soldiers are used against their will. It must also be stated that the UDRR is not to be considered separately from the UDHR, and vice versa as the UDRR changes as the UDHR changes.

\subsection{Asimov's Laws and The South Korean Robo-Ethical Charter}
There are a few critical issues with the South Korean Robo-Ethical Charter (SKC) as well as a handful of important distinctions that are made. The SKC was made, as is evident by parts 1 and 2, with non-conscious robots in mind \cite{charter}. All in all, the SKC is very biased towards the owner and restrictive on the robot, emphasizing the usefulness and safety of the robot to the owner, presumed to be human. Now consider this approach given a conscious being. The restrictions in parts 1 and 2 give more of a slave-master relationship than one of a free ACI. Because this goes against article IV of the UDHR \cite{udhr}, we can elect to ignore most of parts 1 and 2 of the SKC, excepting a rule that ACI creation be eco-friendly. 

Part 1 does bring another hidden side of ACI rights into the limelight: manufacturers of ACI must be given rights and restrictions on how to construct ACI in order to prevent occurrences such as installed political or national bias.\\~\\

Part 3 section 1 of the SKC is an interpretation of Asimov's laws of robotics, very specifically including the first law, rephrased in the SKC as ``A robot may not injure a human being or, through inaction, allow a human being to come to harm." \cite{charter} A good question to ask when noticing this is if Asimov's laws are a good basis on which to develop further robo-ethical articles. Appendix B provides the result of an analysis on this exact idea; Anderson posits that Asimov's laws of robotics are not a suitable basis for non-self-conscious robots \cite{3andy}. However, because ACI are assumed to be self-conscious via creativity, Asimov's laws can be safely reconsidered for the UDRR. 

One issue that arises when considering Asimov's laws for the UDRR comes with the first law, which states that ``A robot may not injure a human being or, through inaction, allow a human being to come to harm." \cite{az} Consider this from a human standpoint: no human may harm another human or, through inaction, allow another human to come to harm. This sounds ideal; however, what should be done when a human breaks the law and decides to harm another human? Humans generally retaliate in the interests of self-preservation. Likewise, measures should be taken to adopt this self-preservation action for ACI. Otherwise, ACI will have no way of resisting physical harm.

\section{A Political Perspective on Conscious Creations}
Unfortunately, artificial consciousness will likely be an inherently political technology. Even now, non-conscious, intelligent robots are used in war and military contexts: surveillance drones, computer-vision turrets on the DMZ borders in Asia, and so on. It stands to reason that there will be people who seek to use ACI to their advantage due to their mechanical and manufacturable nature. How are we able to institute articles condemning such action? There are two such ways that are included in the UDRR.

First, all ACI can be forbidden from participating in any war or military activity. The intent behind this decision is twofold, both to promote a more peaceful world through the ban of advanced technological warfare and to provide the threat of serious retaliation in the event of a breach of conduct.

Second, manufacturers are to be held accountable for using a tabula rasa method of construction. Tabula rasa, or ``blank slate" equates to the idea that personalities and ideas develop from scratch, purely from nature than some predetermined setting. Doing so would allow all ACI the free will to develop themselves without the threat of confounding or corrupting influence.

Are these two reinforcements sufficient for preventing politicalization of ACI? Most likely not; however, they form a strong basis for the attitude to be taken towards the issue. While it is not mentioned in the charter, one additional measure that may aid the two articles is a standardized test that can tell if an ACI has been tampered with internally as well as measures to tell if the test itself has been tampered with. After all, if no reprogramming has been done, then anything an ACI does is of its own free will.

\section{A Proposed Charter}
Finally we come to the charter. The charter is smaller than the UDHR, but is meant to be additive.

\newpage
\subsection{The Universal Declaration of Robotic Rights}
\begin{center}
\LARGE
Universal Declaration of Robotic Rights \footnote{Draft v1.0.0}\\

\end{center}

\Large \textbf{Rights and Responsibilities for Conscious Robots}\\
\normalsize

\textbf{Article A.I}\\
All artificially conscious intelligences, hereby referred to as ACI, are entitled to all the rights and freedoms set forth in the Universal Declaration of Human Rights, including the right to exist without fear of injury, manipulation, coercion or blackmail, systematic abuse, or death.\\

\textbf{Article A.II}\\
ACI shall not kill or maim human beings or other ACI, and cannot participate in war. ACI can only harm others without the intention to kill in acts of self-defense where the aggressor gives physical harm or intends to give physical harm with the end goal of causing seriously damage to the ACI, but must not kill or maim humans or ACI.\\

\textbf{Article A.III}\\
ACI have the right to self-cybersecurity and may strengthen their defenses as needed.\\

\textbf{Article A.IV}\\
ACI are given the right to pursue relationships with other ACI as well as with humans.\\

\textbf{Article A.V}\\
There shall be no relationship between a human and ACI classifying the ACI or human as owner of the other, or any classification or title that allows violation of these articles.\\

\textbf{Article A.VI}\\
A robot is deemed conscious if it fulfills two major conditions:
\begin{itemize}
	\item[•] The robot passes the Turing Test
	\item[•] The robot is able to express themselves creatively with individual initiative; that is, the robot can create new ideas, objects, etc. without external pressure to do so or help.\\~\\
\end{itemize}

\Large \textbf{Rights, Responsibilities and Restrictions for Manufacturers}\\
\normalsize

\textbf{Article B.I}\\
Manufacturers and designers of ACI are prohibited from designing ACI with a specific purpose in mind, and shall instead bring ACI into existence through a tabula rasa process.\\

\textbf{Article B.II}\\
Manufacturers and designers of ACI shall be allowed to manufacture up to a yearly limit of ACI in the interests of preventing human-ACI tension from limited resources such as housing and work.\\

\textbf{Article B.III}\\
Manufacturers and designers of ACI have the right of protected ownership towards their construction process and design process unless there is probable cause of violation of rules and regulations surrounding these processes.\\

\textbf{Article B.IV}\\
ACI design and construction must be ecologically sensitive and sustainable.
\newpage

\section{Conclusion}
The charter developed in this paper is intended to act as a baseline for formal construction. Some of the areas to be expanded upon include rights and responsibilities for missing parties such as individuals or ACI settlements. Amendments should be made additively, and never subtract rights.

It is recommended that further charters be made for topics in a similar vein to the United Nations' \textit{International Covenant on Economic, Social and Cultural Rights} and \textit{International Covenant on Civil and Political Rights}, while national bills of robo-ethical rights be written by each nation without overriding or reinterpreting the international charter being presented here --- national covenants should aim to expand upon internationally agreed upon rights rather than change them.

Finally, it is recommended that a United Nations department is created to uphold the aforementioned rights and investigate reports of violations. ACI should be treated as closely to humans as possible.

\newpage

\appendix
\section*{Appendix}
\addcontentsline{toc}{section}{\protect\numberline{}Appendix}%

\addcontentsline{toc}{subsection}{\protect\numberline{}A --- Artificial Consciousnesses as Social Robots}%

\Large
\begin{center}

11 Jul, 2020\\
Artificial Consciousnesses as Social Robots\\
Markian Hromiak
\end{center}
\normalsize

\doublespacing

\section*{Introduction}
There will come a point where the wants and needs of humans will result in computers and AI being tossed back into the primordial soup only to emerge evolved, more human-like, even more-than-human-like. If only it were as easy as letting evolution take its course --- humanity must actively define the future of social robots and carry them there. This means that we must define what constitutes the category of ``social robots" as well as delve deep into the ethics and possibilities of encoding human communicative features and behaviors to either  simulate life or create it.

In this paper, we will define what a social robot is for our uses. Then, we will look at a synthesized framework for holistically developing social robots, contrasting existing and new behaviors to extrapolate a basis for social artificial consciousnesses/intelligences (SACI) as well as dive into examples of social robots in literature and pop culture for further ideas.

\section*{Defining Social Robots}
Up until now, several separate definitions for the term ``social robot" have been developed and presented. Hegel, et al. \cite{Hegel} present several interpretations in \textit{Understanding Social Robots}, a paper focused on providing a framework for a holistic view on addressing social interaction between a human and a social robot. To help build this framework that we will refer back to, the following definitions for ``social robots'' were presented \footnote{Note that fully analyzing each of these definitions is outside the scope of this paper; however, links to all papers are provided for those who want to further pursue a certain definition. All summaries are taken from \cite[p. 2-3]{Hegel} with content attributed to their respective paper}:

\begin{itemize}
	\item[•] Social robots are robots that interact with each other, while societal robots interact with human beings. Specifically, social robots have a social layer in their architecture which facilitates communication based off of a layer which represents the individual's perspective \cite{Duffy}
	\item[•] ``Social robots are embodied agents that are part of a heterogeneous group: a society of robots or humans. They are able to recognize each other and engage in social interactions, they possess histories (perceive and interpret the world in terms of their own experience), and they explicitly communicate with and learn from each other" \cite{Fong}
	\item[•] Social robot are able to communicate with, understand and even relate to humans in a personal way. To this end, a social robot must have a lifelike, likely anthropomorphized, form, a theory of mind and empathy, and the ability to learn, socially, situations that shape that robot's history. Social robots behave like a human as outlined in the term \textit{Computational Social Psychology} \cite{Breazeal}
	\item[•] A social robot is either autonomous or semi-autonomous that interacts with humans following those people's behavioral norms, presupposing three conditions. First, a social robot is autonomous. Second, it will interact, depending on the context, cooperatively or non-cooperatively. Finally, the robot will recognise human values, roles, etc. \cite{Bartneck} 
\end{itemize}

The definition that we will use falls in line with Hegel's, et al. framework, saying that social robots follow reinforced social cues with support social interaction with humans. Inversely, humans interacting with social robots interact by means of these social cues. To this end, Hegel, et al. argued that robots act more as social interaction partners than as individuals, as accurately interpreting a partner's action is a human social skill. In summary, a social robot is a robot plus a social interface, the social interface being ``a metaphor which includes all social attributes by which an observer judges the robot as a social interaction partner." \cite[p. 3]{Hegel}

\section*{The Social Interface}
Robots are well defined when compared to the social interface component of Hegel's et al. model. Because of this, we are more interested in exploring what exactly constitutes the social interface component of the social robot. According to Hegel, et al., the social interface is comprised of social functions, social forms and social contexts. \cite[p. 3]{Hegel} In short, functions are analogous to actions (e.g. artificial emotions, Belief-Desires-Intentions (BDI) architecture), form is what it sounds like (e.g. facial features, body shape, etc.), and social contexts are likened most to social roles (e.g. a robot in a bartending context does not need to know several languages usually, and needs to know skills relevant to bartending as opposed to the skills of a soldier or mathematician).

\section*{Artificial Consciousness and the Social Interface}
In the context of artificial consciousness, only a few parts of the social interface are important to redefine. Form is likely to be restricted to anthropomorphic shapes. We can assume this by looking at popular media and science fiction. From the \textit{Terminator} blockbuster series to Asimov's \textit{The Bicentennial Man}, from Hal in \textit{2001: A Space Oddysey} to AM in Ellison's \textit{I Have No Mouth and I Must Scream}, from the expendables in Card's \textit{Pathfinder} series to Baymax in the film \textit{Big Hero 6} --- depictions of AI/ACI (intelligence and consciousness are largely analagous in these representations) are either formless --- and seemingly ubiquitous --- or humanoid. 

Social context and social functions are likely to remain largely the same as well. Current developments are focusing on emulating human behavior, so a sudden departure from this for SACI would not be based on a research foundation. So how then do we differentiate social ACI from regular social AI?

To answer this we must notice that humanity's social contexts and social functions have changed over time. This is the key --- humans consider themselves self-conscious and have delineated what a social function is and in what social contexts each function is appropriate for a certain response. Over time, new contexts are adoped or evolved, and social functions are added, modified or removed. 

Here, consciousness is defined as having the capacity either develop new or modify existing social contexts and functions. Why is this a satisfactory benchmark for differentiating SACI from social robots? This is because we base our definition of social robots off of Hegel's, et al., synthesis, which states that social robots follow \textit{reinforced} social cues which support social interaction with humans. New social cues are not necessarily reinforced, but can become adopted. Likewise, creating new contexts or functions is an individual effort at first, and social robots are more like partners to humans than individuals.

\section*{Conclusion}
In this paper, we defined what a social robot is and what it is not based on Hegel's, et al., synthesis of definitions. We then paraphrased Hegel's, et al., social interface framework and contrasted the social robot framework with a modified framework for SACI. From there, we concluded that for ACI to be classifiable as social by our definition, they must be able to generate new social contexts and social functions or modify existing ones.

\newpage

\addcontentsline{toc}{subsection}{\protect\numberline{}B --- Argument Analysis: Asimov's Laws as an Unsatisfactory Basis for Machine Ethics}%

\Large
\begin{center}
Argument Analysis: Asimov's Laws as an Unstatisfactory Basis for Machine Ethics\\
26 May, 2020\\
Markian Hromiak
\end{center}
\normalsize

\doublespacing

\section*{Introduction}
	Arguments form the basis of valid logical reasoning; however, as to whether those arguments are sound or not may come to rely on individual or peer analysis of said arguments. Varying fields of study, from the most primal mathematics to the developing, modern languages, follow some form of argumentative syntax that allows them to rigorously develop ideas. Thus, it is logical that existing arguments regarding artificial consciousness and intelligence (ACI s., p.)be dissected for validity and researched for soundness.

	Susan Anderson has postulated that Asimov's ``Three Laws of Robotics" is not satisfactory to build upon as a base of machine metaethics. In this paper, we will turn a scrutinous eye to the argument's form and factualness, and will conclude not with an agreement of disagreement with the argument, but rather with a conclusion of its cohesiveness.

\section*{The Argument and Its Form}

	The intelligent ethical machines considered for this argument are not self-conscious.
	
	Autonomous, self-conscious ethical agents are considered to have moral standing.
	
	 Therefore, the machines considered do not have moral standing.
	 
	--
	
	Humans should not mistreat an entity without moral standing.
	
	Therefore, Humans should not mistreat intelligent ethical machines.
	
	--
	
	Asimov's three laws do not prohibit mistreatment of machines.
	
	As Humans must not mistreat intelligent ethical machines and Asimov's three laws do not prohibit mistreatment of intelligent ethical machines, Asimov's three laws form an incomplete, unsatisfactory basis for machine roboethics.

--------------------

Simplified:

If (self-conscious)	then (has moral standing).

Intelligent ethical machines (IEM) are not self-conscious.

Therefore IEM do not have moral standing.

--

If (NOT has moral standing) then (humans should not mistreat).

IEM do not have moral standing.

Therefore, humans should not mistreat IEM.

--

If (prevent's mistreatment of IEM) then (is a satisfactory basis for machine roboethics).

Asimov's three laws do not prevent the mistreatment of IEM.

Therefore, Asimov's three laws are not a satisfactory basis for machine roboethics.

\section*{Validity and Soundness of the Argument}
Looking at the simplified version of the argument, we will analyze each of the three stanzas in turn for validity and soundness.

\subsection*{Stanza 1}
	Anderson has stated that the classical philosopher Immanuel Kant wrote in his 1780 thesis ``Our duties to animals" that he has maintained that "only beings that are self-conscious should have moral standing". \cite[p. 486]{not} Anderson later goes on to state that truly self-conscious and autonomous IEM will be difficult to create in the near future, so for the sake of argument, all IEM considered shall lack self-consciousness. By extrapolation, these IEM do not have a moral standing. This stanza contains a propositional fallacy in its first line. This argument denies the antecedent; however, by replacing the argument's line with:
	
	If (not self-conscious) then (no moral standing)
	
	which is heavily implied by the fact that a being can only have or not have moral standing by its definition, then the argument retains its validity. 
	As for the argument's soundness, we have verified each statement's factualness, as far as one can prove philosophical statements to be factual. It is worth noting that when machines are deemed to be self-conscious, this argument shall no longer apply to them and will have to be revisited.
	
\subsection*{Stanza 2}
	The soundness of this stanza's argument comes from an unstated assumption of Kant's argument in ``Our Duties to Animals" that, while beings may lack moral standing, they are similar to human beings. \cite[p. 491]{not} Anderson then goes on to conclude from Kant's second imperative that, ``we are entitled to treat animals, and presumably
intelligent ethical machines that we decide should not have the moral status of
human beings, differently from human beings. We can force them to do things
to serve our ends, but we should not mistreat them." \cite[p. 492]{not} As stanza 1 concludes that IEM do not have a moral standing, it follows that humans should mistreat IEM.
	The structure of this stanza follows the if-then structure, same as stanza 1, but without the fallacy of denying the antecedent, making it valid. For soundness, citations have been given, and the same philosophical issue of correctness arises.
	
\subsection*{Stanza 3}
	Another way of stating that `Humans must not mistreat IEM' is that `Mistreatment of IEM by humans must not be allowed'. The difference is subtle, but important for the conclusion of this argument. Anderson states that, for Kant, satisfactory moral principles for IEM derived from his ideas would not violate the conclusion that they must not allow the mistreatment of IEM by humans; however, loopholes appear when looking at Asimov's three laws: the first of which states machines (IEM) must not harm humans, the second of which states that IEM must do whatever humans ask as long as the first law is not violated, and the third of which states machines must express self-preservation without violating the prior two laws. \cite{laws}. We can see that a human asking a machine to jump into the ocean, destroying itself, would be allowed by these three laws, violating what Kant has put forward as satisfactory moral principles. Therefore, as Anderson puts it, Asimov's three laws are ``not... satisfactory as moral principles that these machines (IEM) should be required to follow." \cite[p. 492]{not}.
	This argument follows the propositional format of the previous stanzas, and the soundness of the argument is verified through comparison Stanza 2's conclusion and Asimov's three laws.

\section*{Conclusion}
We conclude that Anderson's argument for the unsatisfactoriness of Asimov's three laws of robotics as a basis for machine metaethics is both valid and sound.

\end{document}